\documentclass[twocolumn,showpacs,preprintnumbers,amssymb,amsmath]{revtex4}
\usepackage{graphicx}% Include figure files
\usepackage{dcolumn}% Align table columns on decimal point
\usepackage{bm}% bold math

\newcommand{\bi}{\begin{itemize}}
\newcommand{\ei}{\end{itemize}}
\newcommand{\be}{\begin{equation}}
\newcommand{\ee}{\end{equation}}
\newcommand{\ba}{\begin{eqnarray}}
\newcommand{\ea}{\end{eqnarray}}
\newcommand{\bse}{\begin{subequations}}
\newcommand{\ese}{\end{subequations}}
\newcommand{\M}{{\cal {M}}}
\newcommand{\F}{{\cal {F}}}

\newcommand{\Q}{{\cal {Q}}}
\newcommand{\V}{{\cal {V}}}
\newcommand{\C}{{\cal {C}}}
\newcommand{\CP}{{\cal {P}}}
\newcommand{\la}{\langle}
\newcommand{\ra}{\rangle}
\newcommand{\kB}{k_{_B}}

\newcommand{\rvir}{r_{\textrm{\tiny{vir}}}}

\newcommand{\Mvir}{M_{\textrm{\tiny{vir}}}}

\newcommand{\rhoNFW}{\rho_{\textrm{\tiny{NFW}}}}
\newcommand{\MNFW}{M_{\textrm{\tiny{NFW}}}}
\newcommand{\PhNFW}{\Phi_{\textrm{\tiny{NFW}}}}
\newcommand{\VNFW}{V_{\textrm{\tiny{NFW}}}}
\newcommand{\dd}{\textrm{d}}

\begin{document}

\preprint{APS/123-QED}

\title{A general relativistic approach to the Navarro--Frenk--White galactic
halos.} 

\author{Tonatiuh Matos$^\dagger$, Dar\'\i o N\'u\~nez$^\ddagger$ and 
Roberto A Sussman$^\ddagger$}

%\email{sussman@nuclecu.unam.mx}

\affiliation{ 
$^\dagger$ Departamento de F{\'\i}sica,
Centro de Investigaci\'on y de Estudios Avanzados del IPN, A.P. 14-740,
07000 M\'exico D.F., M\'exico.\\$^\ddagger$Instituto de Ciencias Nucleares  
Universidad Nacional Aut\'onoma de M\'exico \\
A. P. 70-543,  M\'exico 04510 D.F., M\'exico\\}

%\date{\today}

\begin{abstract}
Although galactic dark matter halos are basically Newtonian structures,
the study of their interplay with large scale cosmic evolution and with
relativistic effects, such as gravitational lenses, quintessence sources or
gravitational waves, makes it necessary to obtain adequate relativistic
descriptions for these self--gravitating systems. With this purpose in mind, we
construct a post--Newtonian  fluid framework for the
``Navarro--Frenk--White'' (NFW) models of galactic halos that follow from N--body
numerical simulations. Since these simulations are unable to resolve
regions very near the halo center, the extrapolation of the fitting formula leads
to a spherically averaged ``universal'' density profile that diverges at the
origin. We remove this inconvenient feature by replacing a small central region of
the NFW halo with an interior Schwarzschild solution with constant density,
continuously matched to the remaining NFW spacetime. A model of a single halo, as
an isolated object with finite mass, follows by  smoothly matching the NFW
spacetime to a Schwarzschild vacuum exterior along the virial radius, the
physical ``cut--off'' customarily imposed, as the mass associated with NFW
profiles diverges asymptotically. Numerical simulations assume weakly interacting
collisionless particles, hence we suggest that NFW halos approximately satisfy an
``ideal gas'' type of equation of state, where mass--density is the dominant
rest--mass contribution to matter--energy, with the internal energy contribution
associated with an anisotropic kinetic pressure. We show that, outside the
central core, this pressure and the mass density roughly satisfy a polytropic
relation. Since stellar polytropes are the equilibrium configurations in Tsallis'
non--extensive formalism of Statistical Mechanics, we argue that NFW halos might
provide a rough empirical estimate of the free parameter $q$ of Tsallis'
formalism.            
\end{abstract}


\maketitle

\section{Introduction}

A large amount of compelling evidence based on direct and indirect
observations: rotation velocity profiles, microlensing and tidal effects
affecting satellite galaxies and galaxies within galaxy clusters, reveals that
most of the matter content of galactic systems is made up of dark matter (DM).
Since the physical nature of DM so far remains uncertain, this issue has become
one of the most interesting open problems in astrophysics and cosmology
\cite{KoTu,Padma1,Peac,Ellis,Fornengo}. Among a wide variety of
proposed explanations we have: thermal sources, meaning a colissionless gas of
weakly interacting massive particles (WIMP's), which can be very massive ($m\sim
100-200$ GeV) and supersymmetric~\cite{Ellis} (``cold dark matter'' CDM) or
self--interacting less massive ($m\sim$ KeV) particles~\cite{scdm,wdm} (``warm''
DM). 

The DM contribution dispersed in galactic halos is about 90--95
\% of the matter content of galactic systems, while visible baryonic matter
(stars and gas) is clustered in galactic disks. It is then a good
approximation to consider the gravitational field of a galaxy as that of its
DM halo (for whatever assumptions we might make on its physical nature),
while visible matter can be thought of as ``test particles'' in this
field~\cite{HGDM, Lake}.

Assuming the CDM paradigm, we can distinguish two types of halo models:
idealized models obtained from a Kinetic Theory approach, whether based
on specific theoretical considerations or on convenient ansatzes that fix a
distribution function satisfying Vlassov's equation~\cite{BT}, or models
based on ``universal'' mass density profiles obtained empirically from the
outcome of N--body numerical simulations~\cite{nbody_1,nbody_2,nbody_3}. In this
paper we will study the equilibrium configurations that emerge from the latter
approach, based on the well known numerical simulations of Navarro, Frenk and White
(NFW)~\cite{nbody_1,LoMa}. It is important to mention that these
simulations yield virialized equilibrium structures that reasonably fit CDM
structures at a cosmological scale ($\agt 100$ Mpc), though some of their
predictions in smaller scales (``cuspy'' density profiles and excess
substructure) seem to be at odds with
observations~\cite{cdm_problems_1,cdm_problems_2}, especially those based on
galaxies with low surface brightness (LSB), which are supposed to be
overwhelmingly dominated by DM and so well suited to examine the predictions
of various DM models~\cite{vera,LSB}. 

We consider in this paper that DM halos are spherically symmetric
equilibrium configurations, a reasonable approximation since their global
rotation is not dynamically significant~\cite{urc}. 
Galactic halos in virialized equilibrium are also Newtonian systems
characterized by typical velocities, ranging from a few km/sec for dwarf
galaxies up to about $1000-3000$ km/sec for rich clusters. So, why bother
with a general relativistic treatment? First, there is a purely theoretical
interest in incorporating these important self--gravitating systems into
General Relativity, the best available gravitational theory. In fact,
important experimental tests of General Relativity are currently and customarily
carried on (within a weak field post--Newtonian approach) in Solar System bound
Newtonian systems. Secondly, a post--Newtonian description of galactic halos can
be, not only useful and interesting, but essential for studying their interaction
with physical effects that lack a Newtonian equivalent, such as gravitational
lenses or gravitational waves. In fact, the post--Newtonian halo models that we
present in this paper can be readily used in lensing studies, or can provide the
unperturbed zero order configuration in the examination of the perturbative effect
of gravitational waves on galactic halos. Also, a post--Newtonian description
is necessary in the study of the interplay between galactic structures and large
scale ($> 100\,$ Mpc) cosmological evolution dominated by a repulsive ``dark
energy'', modeled by sources like quintessence and/or a cosmological
constant, whose Newtonian description might be inadequate. Finaly,  since galactic
halos are customarily examined as Newtonian structures, we feel it is important to
show the readership of General Relativity journals how to construct spacetimes,
in a post--Newtonian approximation, that are suitable for the description and
study of these important self--gravitating systems. 

 Given the NFW mass density profile, we show in section II how 
all dynamical variables of the Newtonian  NFW halo can be derived. In
section III we construct a post--Newtonian fluid relativistic generalization
of a NFW halo, under the assumption that the gas of collisionless WIMPs should
satisfy an ``ideal gas'' type of equation of state~\cite{RKT,Padma2,Padma3}.
For isotropic velocity distributions, this assumption allows us to determine the
internal energy density by means of the hydrostatic equations themselves. For the
case with anisotropic velocities we follow the same
procedure with regards to the ``radial'' component of the stress tensor,
determining the ``tangential'' stress by a suitable empirical ansatz (section
VI--B).

The fact that the NFW spherically averaged mass density profile diverges at the
halo center follows from interpolating a fitting formula associated with 
numerical simulations that have a finite resolution limit at the
halo center~\cite{NSres}. Although this behavior of the density profile 
does not imply that simulations predict an infinite central density, it is
none--the--less an undesirable feature, which we ammend in 
section IV  by replacing a small region around the center of the NFW halo with a
spherical section of a spacetime with constant matter--energy density, {\it i.e.}
a section of the ``interior'' Schwarzschild solution, that is continuously
matched to the remaining of the NFW post--Newtonian spacetime.
 
Galactic halos are hierarchical structures: small halos lie within
galaxy clusters, which might be part of superclusters, etc, the asymptotic
field of a typical NFW halo should somehow merge with a mean field of a larger
substructure, or with a mean cosmological field. However, the dynamical input
from a suitable cosmological background is well imprinted in the theoretical
design of NFW simulations~\cite{nbody_1,nbody_2,nbody_3}, while the effect of a
cosmological constant on the equilibrium of virialized halo structures is know to
be negligible (see
\cite{SussHdez} and references quoted therein).  Hence, at galactic scales the
empiric NFW profiles are assumed to be valid only up to the ``virial radius'', a
physical ``cut--off'' scale associated with a virialization
process~\cite{Padma1,Peac,BT,Padma2,Padma3}, ignoring altogether their transition
to background fields associated with larger structures or to a cosmological
background. While a post--Newtonian approach also allows one to impose this
virial cut--off scale and to ignore its asymptotic behavior, we show in section
IV how well behaved asymptotically flat NFW configurations can be constructed.
Also, since any localized self--gravitating system (even if belonging to large
substructures) can be approximately described as an isolated system, we also
examine an alternative cut--off by matching generic NFW halos to a Schwarzschild
vacuum exterior at the virial radius.

In section V we provide the equilibrium equations given in terms of suitable
dimensionless variables for the post--Newtonian NFW halos, using the matching
with the ``interior'' and ``exterior'' Schwarzschild solutions defined in
section IV. Analytic solutions of these equations are obtained in section
VI, for isotropic velocities (subsection A) and for a well defined case with
anisotropic velocities (subsection B). We show in both cases that (outside
the central region) the radial pressure and mass density satisfy approximately a
polytropic relation characteristic of stellar polytropes~\cite{BT}. Even if
NFW halos exhibit (in general) deviations from an isotropic velocity
distribution, while velocities in polytropes are strictly isotropic, we argue in
section VII that the resemblance of outer regions of NFW halos to stellar
polytropes might be significant, since virialized self--gravitating systems
exhibit non--extensive forms of energy and entropy, and stelar polytropes are
the equilibrium states in the application to astrophysical systems of the
non--extensive Statistical Mechanics formalism developed by
Tsallis~\cite{Tsallis,PL,TS1,TS2} (see \cite{Chavanis} for a critical approach to
this formalism).

\section{The NFW dark matter halos.}

The well known N--body numerical simulations by
Navarro, Frenk and White (NFW) yield the following ``universal''
expression for the density profile of virialized galactic halo
structures~\cite{nbody_1,nbody_2,nbody_3,LoMa}
\begin{equation}
\rho_{_{\mathrm{NFW}}}=\frac{\delta_0\,\rho_0}{x\,\left(1+x
\right)^2},
\label{rho_NFW}
\end{equation}
where
\begin{eqnarray}
x &=& \ \frac{r}{r_s},\qquad r_s \ = \ \frac{\rvir}{c_0},\label{x} \\
\rho_0 &=& \ \rho_{\mathrm{crit}}\,\Omega_0\,h^2 \ = \ 253.8 \,
h^2\,\Omega_0\,
\frac{M_\odot}{\text{kpc}^3},  \label{rho0} \\
\delta_0 &=& \ \frac{\Delta\,c_0^3}{3\left[\ln\,(1+c_0)-c_0/
(1+c_0)\right]},
\label{delta0} 
\end{eqnarray}
while the concentration parameter $c_0$ can be expressed in terms of
the virial mass $\Mvir$ by~\cite{c0}
\begin{equation}
c_0 = 62.1 \times
\left(\frac{\Mvir\,h}{M_\odot}\right)^{-0.06}\,\left(1+\epsilon\right),
\label{c0}
\end{equation}
where $-1/3\alt \epsilon\alt 1/2 $. The virial radius $\rvir$ is given in
terms of $\Mvir$ by the condition that average halo
density equals $\Delta$ times the cosmological density $\rho_0$
\begin{equation}
\Delta\, \rho_0 = \frac{3\, \Mvir}{4\,\pi\,\rvir^3},  \label{rvir}
\end{equation}
where $\Delta$ is a model--dependent numerical factor (for a $\Lambda$CDM
model with total $\Omega_0=1$ we have $\Delta\sim 100$~\cite{LoHo}). 
Hence all quantities depend on a single free parameter $\Mvir$ with a
dispersion range given by $\epsilon$ for different halo concentrations. 

Using this profile, the mass function and gravitational potential follow from the
Newtonian equations  of hydrostatic equilibrium
\ba M\,' = 4\,\pi\,\rho\,r^2,\label{Mr}\\
\Phi\,' = \frac{G M}{r^2},\label{Phir}\ea
where a prime denotes derivative with respect to $r$. Hence, the NFW
mass function follows from integrating (\ref{Mr}) for
$\rho$ given by (\ref{rho_NFW}) 
\begin{eqnarray}
\MNFW =
4\,\pi\,r_s^3\,\delta_0\,\rho_0\,\left[\ln(1+x)
-\frac{x}{1+x}\right],  \label{M_NFW}
\end{eqnarray}
so that $\MNFW(0)=0$, while $\MNFW$ evaluated at $r=\rvir$ \,\, (or
$x=c_0$) yields $ \Mvir$ as defined in (\ref{rvir}) for $\delta_0$ given by
(\ref{delta0}). Circular rotation velocity and the gravitational potential
follow from (\ref{Phir})
\ba \VNFW^2 && = \ V_0^2\,\,\left[\frac{\ln(1+x)}{x}
-\frac{1}{1+x}\right],  \label{V_NFW}\\
\PhNFW && = \
-V_0^2\,\,
\frac{\ln\,(1+x)}{x}, \label{Ph_NFW}\ea
where the characteristic velocity $V_0$ is
\begin{equation}V_0^2 \ = \
4\,\pi\,G\,r_s^2\delta_0\,\rho_0 \ = \ -\PhNFW(0) \ = \
\frac{3\delta_0}{\Delta c_0^2}\,\frac{G
\Mvir}{\rvir},\label{V0}\end{equation}
and the integration constant was chosen so that $\PhNFW\to 0$ as
$x\to\infty$. Notice that, even if $\rhoNFW$ diverges, all
other quantities (but not their gradients) are regular as  as $x\to 0$.

Since numerical simulations usually yield anisotropic velocity 
distributions, we have in general an anisotropic stress tensor so that
``radial'' and ``tangential'' pressures, $P=P_r$ and $P_\perp$ are involved
in the Navier--Stokes equation 
\begin{equation}P\,' = \ -\rho\,\Phi\,'-\frac{2\,\Gamma}{r}\,P,
\label{Pr}\end{equation}
where 
\begin{equation}\Gamma = \frac{P-P_\perp}{P},\label{alpha}
\end{equation}
is the anisotropy factor. Given $\rhoNFW$ and $\MNFW$, the radial and
tangential pressures follow from integrating (\ref{Pr}) for a given choice
of $\Gamma$. For the NFW forms (\ref{rho_NFW}) and (\ref{M_NFW}),
there are analytic solutions of (\ref{Pr}) for $\Gamma=0$ (isotropic case)
and for various empiric forms of $\Gamma$~\cite{LoMa}.

\section{Relativistic generalization}

Under the assumptions that we outlined in the Introduction, the spacetime
metric for an NFW dark matter galactic halo should be a particular case of
the spherically symmetric static line element
\ba \dd s^2 \ = \ -\exp\left(\frac{2\,\Phi}{c^2}\right)\,c^2\,\dd
t^2+\left(1-\frac{2\,G\,M}{c^2\,r}\right)^{-1}\,\dd
r^2\nonumber\\
+r^2(\dd\theta^2+\sin^2\theta\,\,\dd\phi^2),\label{metric}\ea
so that $M(r)$ has units of mass. The functions 
$\Phi(r)$ and $M(r)$ are suitable relativistic generalization of the
NFW functions given by (\ref{M_NFW}) and (\ref{Ph_NFW}). We will assume
a fluid energy--momentum tensor of the most general form for the metric
(\ref{metric}) 
\ba T^{ab}  = \ \mu\,u^a\,u^b + p\,h^{ab} + \Pi^{ab},\label{Tab}\ea
where $\mu$ and $p$ are the matter--energy density and isotropic pressure
along a 4-velocity field $u^a =\exp(-\Phi/c^2)\,\delta^a\,_t$,
\ while $h^{ab}  =  g^{ab}+u^a\,u^b$ and $\Pi^{ab}$ is the anisotropic and
traceless ($\Pi^a\,_a=0$) stress tensor, which for the metric
(\ref{metric}) takes the form
\ba \Pi^a\,_b \ = \ \textbf{diag}\,[0,\,-2\Pi,\,\Pi,\,\Pi]\ea
so that $p$ and $\Pi=\Pi(r)$ relate to the radial and tangential pressures, $P$
and $P_\perp$, by
\begin{equation}P_\perp-P \ = \ 3\,\Pi,\qquad 2\,P_\perp+P \ = \
3\,p.\label{pps}\end{equation}
The field equations and momentum balance ($T^{ab}\,_{;b}=0$) associated with
(\ref{metric})-(\ref{pps}) are
\ba M\,' &&= \ 4\,\pi\,\mu\,r^2/c^2,\label{M2r}\\
\Phi\,' &&= \
\frac{G\,[M+4\,\pi\,P\,r^3/c^2]}{r\,\left[r-2\,G\,M/c^2\right]},
\label{Phi2r}\\ P\,'&& = \
-(\mu+P)\,\frac{\Phi\,'}{c^2}-\frac{2\,\Gamma}{r}\,P,\label{P2r}\ea
where $\Gamma$ is given by (\ref{alpha}). These
equations are the relativistic generalization of the Newtonian equilibrium
equations (\ref{Mr}), (\ref{Phir}) and (\ref{Pr}). In the Newtonian
case all these equations are decoupled, so that once $\rho$ is known and
$\Gamma$ is prescribed, all other quantities follow by simple integration of
quadratures. In the relativistic case we have, in general, three equations for
five unknowns ($\mu,\,P,\,\Gamma,\,M,\,\Phi$). Thus, we must provide a relation
between $\mu$ and $P$, together with a suitable assumption that determines or
prescribes the form of $\Gamma$. 

Since the WIMPs in the collisionless gas making up
galactic halos are interacting very weakly, it is reasonable to consider such
a gas as approximately an ``ideal gas'' whose total matter--energy density,
$\mu$, is the sum of a dominant contribution from rest--mass density,
$\rho\,c^2$, and an internal energy term that is proportional to the pressure
$P$ and to the velocity dispersion $\sigma^2=\la v^2\ra\simeq \la
v_\perp^2\ra$. Hence, we shall assume that the matter source of NFW halos
complies with the equation of state of a non--relativistic (but non--Newtonian)
ideal gas~\cite{Peac,RKT,Padma2,Padma3} 
\begin{equation} \mu \ = \
\rho\,c^2\,\left[1+\frac{3}{2}\,\frac{\sigma^2}{c^2}\right],\qquad P
\ = \ \rho\,\sigma^2,\label{NRIGES}\end{equation}
where we can identify $\rho=\rhoNFW$ (or with any mass density formula used
in halo models) and the velocity dispersion is related to a kinetic
temperature
$T$ by~\cite{BT}
\begin{equation}\sigma^2 \ = \ \frac{P}{\rho} \ = \ 
\frac{\kB\,T}{m},\label{sigma2}\end{equation}     
where $\kB$ is Boltzmann's constant. At this point, we believe it is convenient
to mention the following two idealized models of self--gravitating systems as
useful theoretical references~\cite{BT}:
\ba &&\textrm{Isothermal Sphere:}\nonumber\\&&\sigma^2 = \frac{\kB\,T_0}{m}
=\textrm{const.}  \quad \Rightarrow\quad P = K\,\rho,\qquad
\nonumber\\
\nonumber\\
&&\textrm{Stellar Polytropes:}\nonumber
\\
&&\sigma^2 = K\,\rho^{1/n} \quad\quad\, \Rightarrow\quad P =
K\,\rho^{1+1/n},\qquad
\label{isotpoly}\ea
where $T_0$, $K$ and $n$ (polytropic index) are constants. The isothermal sphere 
corresponds to a Maxwell--Boltzmann velocity distribution, the equilibrium state
associated with the extensive Boltzmann--Gibbs entropy~\cite{BT,Padma2,Padma3}. 
The stellar polytropes are also solutions of the Vlassov equation, but are
associated with the equilibrium state in the non--extensive entropy functional
proposed by Tsallis~\cite{Tsallis,PL,TS1,TS2}. Notice that the isothermal sphere
follows from the stellar polytropes in the limit $n\to\infty$ (the extensivity
limit in Tsallis' formalism).     
  
For Newtonian characteristic velocities in galactic halos, we have
$\sigma^2/c^2\ll 1$ and $\mu\approx \rho\,c^2$ and so $P\simeq P_\perp\ll
\rho\,c^2 $, so that (\ref{NRIGES}) provides a plausible equation of state
for a relativistic generalization of galactic halos. It is evident 
that in the Newtonian limit $\sigma^2/c^2\to 0$ we recover the Newtonian
equilibrium equations (\ref{Mr}), (\ref{Phir}) and (\ref{Pr}). 
What needs to be done now is to insert the equation of state (\ref{NRIGES})
into the field equations (\ref{M2r})--(\ref{P2r}). It turns out to be
easier to work with $P$ instead of
$\sigma$ or $T$, using (\ref{NRIGES}) as
\begin{equation} \mu \ = \ \rho\,c^2 +
\frac{3}{2}\,P,\label{NRIGES2}\end{equation}
so that $\sigma$ and/or $T$ can be obtained afterwards from $P$ through
(\ref{NRIGES}) and (\ref{sigma2}). Combining (\ref{Phi2r}) and
(\ref{P2r}) into a single equation and using (\ref{NRIGES2}) we obtain the
set 
\ba M' &=& \ 4\,\pi\,\left[\rho+\frac{3}{2}\,\frac{P}{c^2}
\right]\,r^2,\label{M3r}\\
P' &=& \
-\frac{G[\rho+\frac{5}{2}P/c^2]\,[M+4\pi\,(P/c^2)\,r^3]}
{r[r-(2G/c^2)\,M]}-\frac{2\,\Gamma}{r}\,P,\nonumber\\
\label{P3r}\ea
which becomes determined once we identify $\rho=\rhoNFW$ and specify 
$\Gamma=\Gamma(r)$. We can solve these equations in a post--Newtonian
approximation by keeping only terms up to order $\sigma^2/c^2$.

\section{Providing a regular center and matching with a
Schwarzschild exterior}

By looking at (\ref{rho_NFW}), it is evident that $\rho=\rhoNFW$ diverges as
$r\to 0$. A careless examination, from a full general relativistic
point of view, of the spherical spacetime given by (\ref{metric})--(\ref{NRIGES})
with $\rho=\rhoNFW$, would yield a curvature singularity marked by $r=
0$, associated with the blowing up of the Ricci scalar
\begin{equation}R \ = \ \frac{8\pi
G}{c^4}\,\left[\rhoNFW\,c^2+\left(2\,\Gamma-\frac{3}{2}\right)\,P\right].
\label{Ricci1}
\end{equation}
However, this situation does not apply to NFW halos, not only because they are
Newtonian systems that must be examined within the framework of a Newtonian limit
of a weak field approach, but because the NFW mass density profile
(\ref{rho_NFW}) is an empirical fitting formula that emerges from
spherically averaging numerical simulations that cannot resolve distances smaller
than about 1 \% of the actual physical radius of the halo~\cite{NSres}. Hence, 
astrophysicists using this density profile do not actually assume infinite
central densities, but regard this blowing up of
$\rhoNFW$ as an undesired effect due to the extrapolation of a fitting formula
which (within the resolution limits of numerical simulations) provides a rough
illustration of the fact that density becomes ``cuspy'' along the central halo
region, {\textit i.e. } $\rhoNFW\sim 1/x$ for $x\ll 1$.   

A practical way to get rid of this inconvenient feature  is to ``replace'' a
small spherical region $0<x<x_0$ of the NFW spacetime with an ``inner'' fluid
region  containing the world--line of a regular center. Using the definitions
(\ref{x})--(\ref{rvir}), the radius $r_0$ of this inner region  in terms of
$\Mvir$ is given by 
\begin{equation}\frac{r_0}{\textrm{kpc}} \ = \
0.272\,\times\,\left(\frac{\Mvir}{M_\odot}\right)^{0.273}\,x_0.
\end{equation}
Hence, for halos in the observed range $10^8\,M_\odot < \Mvir< 10^{15}\,M_\odot$,
the choice $x_0=0.0001$ yields $4\,\textrm{pc}\alt r_0\alt 340\,\textrm{pc}$, a
very small radius in relation to the virial radii of these halos. Thus, since
this length scale is much smaller than the maximal resolution of numerical
simulations~\cite{NSres}, we are able to provide a regular center for
the NFW spacetime but this does not prevent us from studying
the effects of its steep density profile in the central region.

The simplest choice of a spacetime geometry for the inner region is a
section of a Schwarzschild interior solution~\cite{Steph} characterized by
the metric (\ref{metric}) with 
\ba \exp\,\left(\frac{\Phi}{c^2}\right) &=&
a_0-b_0\,\sqrt{1-\kappa_0\,r^2}\nonumber
\\  M &=& \frac{4\,\pi\,\mu_c}{3\,c^2}\,r^3,\label{ISmetric}\ea
with $\kappa_0=8\,\pi\, G\,\mu_c/c^4 $ and
\ba \mu &=& \mu_c \ = \ \textrm{const.},\\
P &=&
\frac{\mu_c}{3}\,\frac{3\,b_0\,\sqrt{1-\kappa_0\,r^2}-a_0}
{a_0-b_0\,\sqrt{1-\kappa_0\,r^2}},\label{P_IS}\ea
where the constants $a_0,\,b_0$ and $\mu_c$ must be selected so that this region can be
suitably ``glued'' to the NFW spacetime occupying $x>x_0$.  

As we mentioned before, it is customary to disregard the asymptotic behavior of
NFW profiles, since the virial radius is considered to be the physical cut--off
radius of NFW halos. However, we can construct asymptotically well behaved 
NFW configurations for which $\mu,\,P,\,M/r$ and $\Phi$ tend to zero as $x\to
\infty$ (though $M$ most certainly will diverge in this limit, since the
Newtonian $\MNFW$ in  (\ref{M_NFW}) already does). A finite $M$ as $r\to \infty$
can be achieved if we match the NFW spacetime at a convenient cut--off scale
to a Schwarzschild vacuum exterior characterized by $\mu=P=0$ and by
(\ref{metric}) with
\ba \exp\,\left(\frac{2\,\Phi}{c^2}\right) \ = \ 
1-\frac{2\,G\,M_0}{c^2\,r}, \qquad M \ = \ M_0,
\label{Smetric}\ea
where $M_0$ is the constant ``Schwarzschild mass''.

Necessary conditions for a smooth matching between spacetime regions are given by
Darmois matching conditions~\cite{seno}, requiring continuity of the induced
metric and extrinsic curvature of the matching hypersurface 
\begin{equation} h_{ab} = n_a\,n_b-g_{ab},\qquad K_{ab} =
-h_a^c\,h_b^d\,n_{c;d},\label{darmois}\end{equation}
where $n^a$ is a unit vector normal to this hypersurface. Since the NFW spacetime 
must be matched, either to (\ref{ISmetric}) or to (\ref{Smetric}), at 
hypersurfaces marked by constant $r$, we have $n_a = \sqrt{g_{rr}}\,\delta_a^r$,
hence (\ref{darmois}) imply that $g_{tt},\,g_{tt}'$ and $g_{rr}$ (but not
necessarily $g_{rr}'$) must be continuous at the matching hypersurface.
Considering (\ref{M2r})--(\ref{P2r}), this implies continuity at the matching
hypersurface of
$M,\,\Phi$ and $P$, but not of $\mu$ or the anisotropic pressure defined in
terms of $\Gamma$ by (\ref{alpha}).

\subsection{Matching with the inner region.}

It is convenient to assume (\ref{NRIGES2}) to be valid at $r=0$, so that we
can characterize the Schwarzschild interior solution by 
\begin{equation} \mu_c \ = \ \rho_c\,c^2+(3/2)\,P_c, \label{mu_intS}
\end{equation}
Following (\ref{V0}), we can define a characteristic velocity
\begin{equation}V_c^2 \ = \ 4\,\pi\,G\,\rho_c\,r_s^2,\label{Vc_intS}\end{equation}
so that
\begin{equation}P_c \ = \ \delta_c\,\rho_c\,V_c^2,\qquad \xi \ = \
\frac{P_c}{\rho_c\,c^2} \ = \
\delta_c\,\frac{V_c^2}{c^2},\label{Pc_intS}\end{equation} 
where $\delta_c$ is an arbitrary constant, so that central velocity dispersion is
$\sigma_c^2=\delta_c\,V_c^2$. Hence, for
$0\leq x \leq x_0$ we have
\begin{equation} 
M \ = \
\frac{4\,\pi}{3}\,\rho_c\,\left(1+\frac{3}{2}\,\xi\right)\,r_s^3\,x^3,
\label{M_intS}\end{equation}
while, for the time being, we assume also $\Gamma=0$, though a nonzero $\Gamma$ can be
considered for the inner region in the case of anisotropic pressure (see
section VI--B). Since we are considering $x<x_0\sim 0.0001 \ll 1$, suitable
expressions for the remaining variables in this region are found by
expressing $a_0,\,b_0$ in terms of the parameters in
(\ref{mu_intS})--(\ref{Vc_intS}) and expanding (\ref{ISmetric}) and
(\ref{P_IS}) up to first order in $x^2$, leading to 
\ba 
\Phi \ &\approx& \
\Phi_c+\frac{1}{6}\,V_c^2\,\left(1+\frac{9}{2}\,
\xi\right)\,x^2,\label{Phi_intS}\\ 
P \ &\approx&
\ \rho_c\,V_c^2\left[\delta_c -
\frac{1}{6}\,\left(1+\frac{5}{2}\,
\xi\right)\,\left(1+\frac{9}{2}\,
\xi\right)\,x^2\right],\nonumber\\
\label{P_intS}\\
V_{\textrm{\tiny{rot}}}^2 &=& r\,\Phi' \ \approx \
\frac{1}{3}\,V_c^2\,\left(1+\frac{3}{2}\,\xi\right)\,x^2,\label{V_intS}\ea
Following the matching conditions (\ref{darmois}), the constants $\rho_c,\,V_c,\Phi_c$
and $\delta_c$ must be selected so that $M(x_0),\,\Phi(x_0)$ and $P(x_0)$ continuously
match the NFW functions $M,\,\Phi$ and $P$ at $x_0$. Although, (\ref{darmois}) do not
require this continuity for $\mu$ and $\Gamma$, we will still assume it in order
to avoid an unphysical jump discontinuity of these variables at $x=x_0$, as well
as all state variables. 

\subsection{Matching with a vacuum exterior.}

A smooth matching with a Schwarzschild exterior at a cut--off radius $r=r_b$ based on
(\ref{darmois}) requires
\ba M(r_b) &=&  M_0,\quad
\textrm{e}^{2\Phi(r_b)/c^2}=1-\frac{2\,G\,M_0}{r_b},\label{darmois31}\\
P(r_b) &=& 0,\label{darmois32}\ea
but do not require $\mu$ or $\Gamma$ to vanish at $r_b$. However, a jump discontinuity
of these variables at an interface with a vacuum exterior is much more acceptable than
in the interface between two non--vacuum regions. As we discuss in section VI, a
convenient cut--off scale for a NFW spacetime is the virial radius $r_b=\rvir$, so
that we can identify $M_0$ with $\Mvir$. Though, because of the matching with the
inner region, an $\rvir$ selected by means of condition
(\ref{darmois32}) will not yield (even in the Newtonian limit)
$M_0=\Mvir$ with $\Mvir$ given by (\ref{rvir}). However, for sufficiently small $x_0\ll
1$ the resulting $M_0$ will be approximately equal to $\Mvir$.  

\section{Post--Newtonian NFW halos}

In order to explore the post--Newtonian limit for the system
(\ref{M3r})--(\ref{P3r}), it is useful to work with
dimensionless variables by rescaling all variables in terms of
quantities defined at the scale radius $r_s$.
 
\subsection{The region $x>x_0$}

Convenient rescalings follow as 
\ba Y \ &=& \ \frac{\rhoNFW}{\delta_0\,\rho_0} \ = \
\frac{1}{x\,[1+x]^2},\label{Y}
\\
\M \ &=& \ \frac{M}{4\,\pi\,\delta_0\,\rho_0\,r_s^3} \ = \
\frac{c_0^3\,\Delta\,M}{3\,\delta_0\,\Mvir},\label{CM}\\
\CP \ &=& \ \frac{P}{\delta_0\,\rho_0\,V_0^2},\label{CP}\\
\Psi \ &=& \ \frac{\Phi-\Phi_c}{V_0^2},\label{Psi}\ea  
with $V_0$ defined in (\ref{V0}), transforming (\ref{M3r}) and (\ref{P3r}) into
\ba \frac{\dd \M}{\dd x}  &&= \
\left[Y+\frac{3}{2}\,\varepsilon\,\CP\right]\,x^2,\label{Mx}\\
\frac{\dd \CP}{\dd x} &&= \
-\frac{\left[Y+\frac{5}{2}\,\varepsilon\,\CP\right]\,\left[\M+\varepsilon\,
\CP\,x^3\right]}{x\,\left[x-2\,\varepsilon\,\M\right]}-
\frac{2\Gamma}{x}\,\CP,\nonumber\\
\label{Px}\ea
where 
\begin{equation} \varepsilon \ = \ \frac{V_0^2}{c^2},
\label{epsilon}\end{equation}
so that in the limit $\varepsilon\to 0 $ we recover the Newtonian 
equations (\ref{Mr}), (\ref{Phir}) and (\ref{Pr}). The system
(\ref{Mx})--(\ref{Px}) can be integrated by demanding that $\M$ and $\CP$ comply
with appropriate boundary and initial conditions, so that the NFW halo can be smoothly
matched with the Schwarzschild interior at $x=x_0$ and the Schwarzschild exterior at
$r=\rvir$. Since we have to use the explicit form of $Y$ in (\ref{Y}), then the
analytic or numerical solutions of (\ref{Mx})--(\ref{Px}) for specific choices of
$\Gamma$, boundary conditions depend on $\Mvir$ through the definitions
(\ref{delta0}) and (\ref{c0}). 

The metric function $M=V_0^2\,r_s\,\M$ follows from (\ref{Mx}), while 
$\Phi=\Phi_c+V_0^2\,\Psi$ can be obtained by integrating
\begin{equation} \frac{\dd\,\Psi}{\dd x} \ = \ 
\frac{\M+\varepsilon\,\CP\,x^3}{x\,[x-2\,\varepsilon\,\M]}.
\end{equation}
The relativistic generalization of the
Newtonian rotation velocity profile are the velocities of test observers
along circular geodesics. These velocities are \cite{HGDM,Lake}
$V_{\textrm{\tiny{rot}}}^2=r\,\Phi'$, which in terms of the dimensionless
variables becomes
\begin{equation}\V^2 \ = \ \frac{V_{\textrm{\tiny{rot}}}^2}{V_0^2} \ = \
\frac{\M+\varepsilon\,\CP\,x^3}{x-2\,\varepsilon\,
\M}\end{equation}

Since $V_0$ for typical galactic halos ranges from a few km/sec to $\sim
1500$ km/sec, the post--Newtonian corrections of order $V_0^2/c^2$ will be
very small: between $O(\varepsilon)\sim 10^{-9}$ and $O(\varepsilon)\sim
10^{-6}$. The post--Newtonian system associated with
(\ref{Mx})--(\ref{Px}) can be given as
\ba \frac{\dd \M}{\dd x}  &=& \
Y\,x^2+O(\varepsilon),\label{Mx0}\\
\frac{\dd \CP}{\dd x} &=& \
-\frac{Y\,\M}{x^2}-
\frac{2\Gamma}{x}\,\CP+O(\varepsilon),
\label{Px0}\ea
with
\ba\frac{\dd\,\Psi}{\dd x} \ = \
\frac{\M}{x^2}+O(\varepsilon),\label{PhiV}\\ 
\V\,^2 \ = \
\frac{\M}{x}+O(\varepsilon) 
\ea

\subsection{The region $0\leq x\leq x_0 $}

The variables defined in the previous subsection must glue continuously at $x_0$
with the interior Schwarzschild variables (\ref{mu_intS})--(\ref{P_intS}).
Normalizing these variables with the same factors as in
(\ref{Y})--(\ref{CP}), we have in the region $x\leq x\leq x_0$
\ba Y &=& Y_0  \ = \ \frac{\rho_c}{\delta_0\,\rho_0} \
= \ \frac{\xi}{\delta_c\,\varepsilon} \ = \ 
\frac{1}{x_0\,(1+x_0)^2},\label{Y0}\\
\M &=&
\frac{1}{3}\,Y_0\,\left[1+\frac{3}{2}\,\xi\right]\,x^3,\\
\CP &\approx&
Y_0^2\,\left[\delta_c-\frac{1}{6}\,\left(1+\frac{5}{2}\,
\xi\right)\,\left(1+\frac{9}{2}\,
\xi\right)\,x^2\right],\label{P_intSPN0}\nonumber\\\ea
\ba
\Psi  &\approx& 
\frac{1}{6}\,Y_0\,
\left(1+\frac{9}{2}\,
\xi\right)\,x^2,\label{Psidef}\\
\V\,^2 &\approx& 
\frac{1}{3}\,Y_0^2\,\left(1+\frac{3}{2}\,\xi\right)\,x^2.\ea
From (\ref{Y0}) and bearing in mind that $x_0\ll1$, we have 
\begin{equation}\xi \ \approx \
\frac{\varepsilon\,\delta_c}{x_0},\label{xieps}\end{equation}
implying that $\xi\sim \varepsilon$ if $\delta_c\sim x_0$. Since
$\varepsilon$ is very small we can also assume that $\xi\ll 1$, so that
post--Newtonian expressions follow by taking only terms up to $O(\xi)$:
\ba \M &=&
\frac{1}{3}\,Y_0\,x^3 + O(\xi),\label{M_intSPN}\\
\CP &=&
Y_0^2\,\left[\delta_c-\frac{1}{6}\,x^2\right]+
O(\xi),\label{P_intSPN}\\
\Psi &=&  \frac{1}{6}\,Y_0\,
x^2+ O(\xi), \label{Phi_intSPN}\\
\V\,^2 &=& 
\frac{1}{3}\,Y_0^2\,x^2+ O(\xi).\label{V_intSPN}\ea

 We examine analytic solutions of the post--Newtonian system
(\ref{Mx0})--(\ref{Px0}) that match continuously with
(\ref{M_intSPN})--(\ref{Phi_intSPN}).

\begin{figure}
\centering
% Use the relevant command for your figure-insertion program
% to insert the figure file.
% For example, with the option graphics use
\includegraphics[height=20cm]{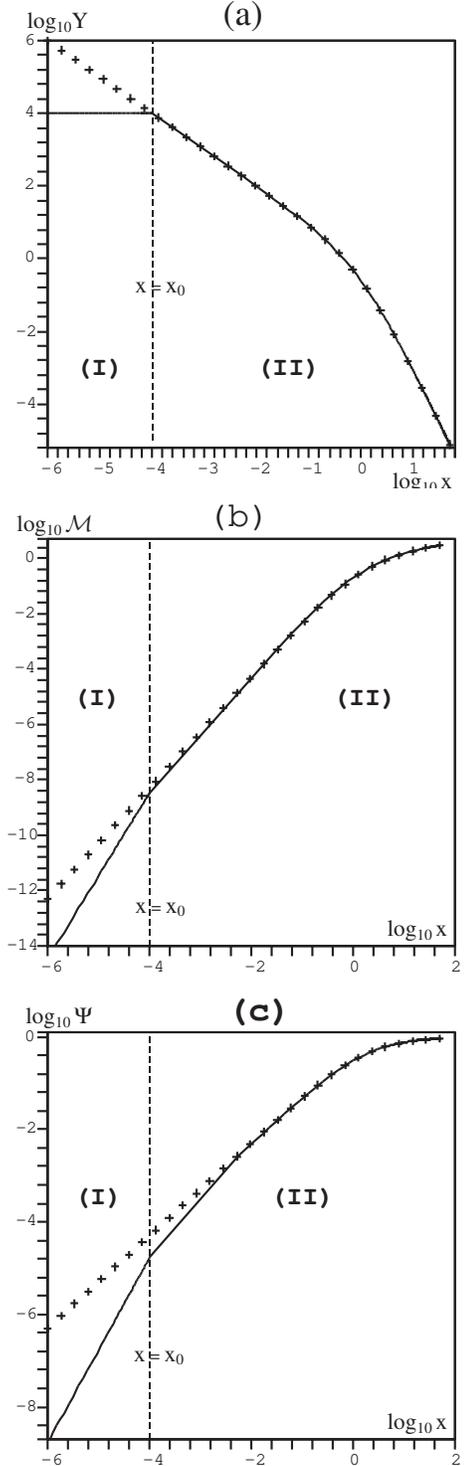}
%
% If not, use
%\picplace{5cm}{2cm} % Give the correct figure height and width in cm
%
\caption{Solid curves in panels (a), (b) and (b) respectively depict the logarithmic
plots of $Y,\,\M$ and $\Psi$ given by (\ref{Y}), (\ref{Y0}), (\ref{eq_M}) and
(\ref{eq_Psi}), in the inner (I) and outer (II) regions separated by $x_0=0.0001$. In
each panel the ``pure'' NFW case without inner region is shown by the curves with
crosses. Notice how this latter case is practically identical to $Y,\,\M$ and $\Psi$ in
the outer region.}
\label{fig:1}       % Give a unique label
\end{figure}

\section{Analytic solutions.}

For whatever choice of $\Gamma$ and restrictions on $P$, equations (\ref{Mx0})
and (\ref{PhiV}) can be integrated, yielding $\M$ and $\Phi$ so that $\M(x_0)$
and $\Phi(x_0)$ match (\ref{M_intSPN}) and (\ref{Phi_intSPN}) at $x=x_0$.
Denoting the inner and NFW regions as
\ba 0 \ \leq \ x \ \leq \ x_0, \qquad \textrm{(I)},\nonumber\\
 x \ > \ x_0, \qquad \textrm{(II)}. \label{zones} \ea
We have then the following solutions up to orders $O(\varepsilon)\sim O(\xi)$: 
\ba \M_{\textrm{(I)}} &=& \frac{x^3}{3\,x_0\,(1+x_0)^2}+O(\xi),\nonumber\\
\M_{\textrm{(II)}}
&=&\frac{(3+4x_0)\,x_0}{3\,(1+x_0)^2}+\ln\,\frac{1+x}{1+x_0}-\frac{x}{1+x}
+O(\varepsilon),\nonumber\\
\label{eq_M}\ea
\ba\Psi_{\textrm{(I)}} &=&
\frac{x^2}{6\,x_0(1+x_0)^2}+O(\xi),\nonumber\\
 \Psi_{\textrm{(II)}} &=&
\frac{2+3\,x_0}{2\,(1+x_0)^2}+\frac{1}{x}\,\left[\ln\,\frac{1+x}{1+x_0}
-\frac{(3+4\,x_0)\,x_0}{3\,(1+x_0)^2}\right]\nonumber\\&&+O(\varepsilon),
\label{eq_Psi}\ea
Therefore, irrespective of the choice of $\Gamma$ and/or assumptions on $P$,
the metric elements for all NFW halo spacetimes are up to order
$\varepsilon^2$
\ba
-g_{tt} &=& 
\textrm{e}^{2\Phi_c/c^2}\textrm{e}^{2\varepsilon\,\Psi}
= 1+\frac{2\Phi_c}{c^2}+2\,\varepsilon\Psi+O(\varepsilon^2),\nonumber\\
\label{gtt}\\
 g_{rr} &=& 
\left[1-\frac{2\,\varepsilon\,\M}{x}\right]^{-1}= 
1+\frac{2\,\varepsilon\,\M}{x}
+O(\varepsilon^2),\label{grr}
\ea 
where the metric functions $\M$ and $\Psi$ are given by (\ref{eq_M}) and (\ref{eq_Psi})
in the regions (I) and (II), and we have obtained $M$ and $\Phi$ from $\M$ and $\Psi$
by means of (\ref{CM}) and (\ref{Psidef}). Figures 1a, 1b and 1c display the
normalized density $Y$ and the metric functions $\M$ and $\Psi$ in regions (I) and
(II).

Notice that all functions defined so far reduce to their Newtonian NFW forms, as given
in section II, in the limits $x_0\to 0$ and $\varepsilon\to 0$. Also, while all NFW
halos have the same rest--mass density $Y$, the form for the pressure depends on the
assumptions one might make about $\Gamma$ and suitable boundary conditions.

\subsection{Isotropic case}

For $\Gamma=0$ we have $P=P_r=P_\perp$ and so pressure is isotropic. In
collisionless systems this implies an isotropic distribution of velocity
dispersion. In this case, (\ref{Px0}) yields the following analytic solution
\ba \CP_{\textrm{(I)}} &=& \frac{\delta_c-x^2/6}{x_0^2\,(1+x_0)^4}+O(\xi)\nonumber\\
\CP_{\textrm{(II)}} &=&
\frac{\delta_c-x_0^2/6}{x_0^2\,(1+x_0)^4}+\F-\F_0+O(\varepsilon),\label{CP_iso1}\ea
where $\F_0$ is the evaluation at $x=x_0$ of the function
\begin{widetext}
\ba \F &=& \F(x_0,x) \ = \
\frac{3}{2}\,\left[\ln\,(1+x)\right]^2+\left[\,A_1-\alpha_0\,\right]\,
\ln\,(1+x)+\left[\alpha_0-7/2\right]\,\ln x +
3\,\textrm{Li}_2\,(1+x)+\alpha_0\,B_1+C_1,\label{CP_iso2}\\
\alpha_0&=&3\ln(1+x_0)+\frac{3(1+x_0)-x_0^2}{(1+x_0)^2},\quad A_1 =
\frac{1-3x+x^2+7x^3}{2x^2(1+x)}\quad
B_1=\frac{1-3x-6x^2}{2x^2(1+x)},\quad
C_1=\frac{1-3x-18x^2-13x^3}{2x^2(1+x)^2}\nonumber  
\ea  
\end{widetext}
and the dilogarithmic function is defined as
\begin{equation}\textrm{Li}_2(y) \ = \
\int_1^y{\frac{\ln t\,\dd t}{1-t}}.\nonumber\end{equation} 
This form of $\CP$ matches continuously the regions (I) and (II). Since the limit
of $\F(x,x_0)$ as $x\to\infty$ depends explicitly on $x_0$ and
$\delta_c$, we need to find appropriate relations $\delta_c=\delta_c(x_0)$ in
order to determine (together with (\ref{eq_M})--(\ref{grr})) the asymptotic behavior
of the NFW spacetime.

\begin{figure}
\centering
% Use the relevant command for your figure-insertion program
% to insert the figure file.
% For example, with the option graphics use
\includegraphics[height=15cm]{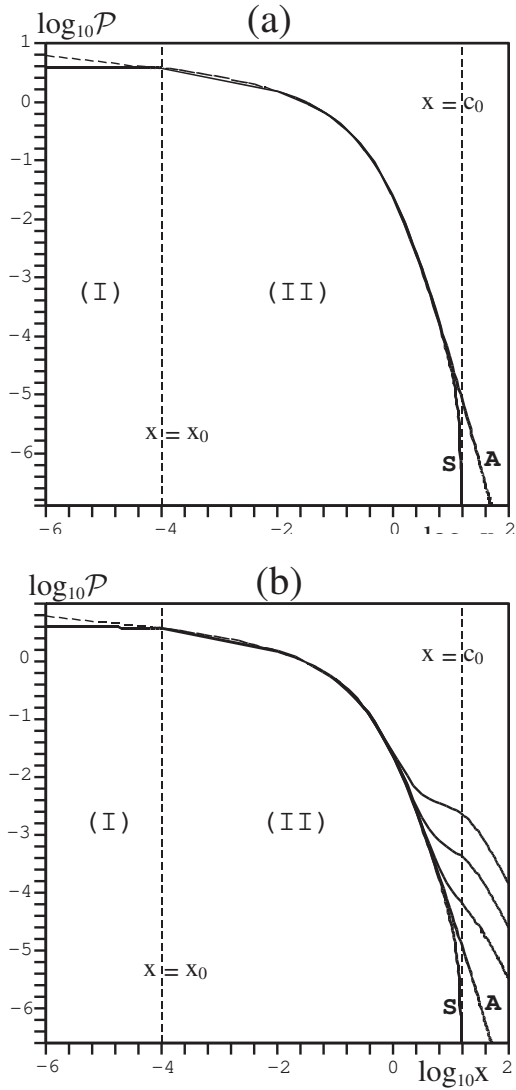}
%
% If not, use
%\picplace{5cm}{2cm} % Give the correct figure height and width in cm
%
\caption{Logarithmic plot of normalized pressure $\CP$ in regions (I)
and (II) with $x_0=0.0001$. Panels (a) and (b) respectively
denote the isotropic (eqs (\ref{CP_iso1})--(\ref{CP_iso2})) and anisotropic (eqs
(\ref{CP_anisI})--(\ref{Qdef})) cases. The letters \textbf{A} and
\textbf{S} depict the asymptotically flat case that scales as \,$\ln\,x/x^4$\, and
the case with a matching with a Schwarzschild exterior at the virial radius
($x=c_0=15$) so that
$\CP(c_0)=0$. The dotted curves in the inner regions are the NFW cases without an
inner region. In the anisotropic case we have chosen $\beta=20$. The curves deviating
from
\textbf{A} correspond to cases that scale asymptotically as $x^{-2}$}
\label{fig:2}       % Give a unique label
\end{figure} 

\subsubsection{Asymptotically flat configuration}

The asymptotic behavior ($x\gg 1$) of (\ref{CP_iso1})--(\ref{CP_iso2}) 
is given by
\ba \CP_{\textrm{(II)}} &=&
\frac{\delta_c-x_0^2/6}{x_0^2\,(1+x_0)^4}-\F_0-\frac{\pi^2}{2}\nonumber\\&+&
\frac{1-(3/4)\,\alpha_0+4\,\ln\,x}{16\,x^4}+O\left(\frac{\ln
x}{x^5}\right),\label{CP_iso_asympt}\ea
Since in region (II) we have: \,$\rho\to 0$ as $x\to\infty$, an asymptotically
flat NFW configuration without cut--off scales requires that $\CP_{\textrm{(II)}}
\to 0$ as $x \to \infty$, thus the zero order term in
(\ref{CP_iso_asympt}) must vanish, leading to
\ba \delta_c &=&
\frac{x_0^2}{6}\left[1+6\,(1+x_0)^4\left(\F_0+\frac{\pi^2}{2}\right)\right],
\nonumber\\
 &\approx& 
\left[\ln\,\frac{1}{\sqrt{x_0}}+\frac{\pi^2}{2}-\frac{17}{3}\right]\,x_0^2 \ \approx \
x_0^2\,\ln\,\frac{1}{\sqrt{x_0}},\label{delta_apr1}
\ea
where we have used (\ref{CP_iso2}) and the fact that $x_0\ll 1$ in order to get this
leading term expansion on $x_0$. Hence, (\ref{xieps}) implies that $\xi$ is smaller
than $\varepsilon$ but of the same order of magnitude, so that $O(\xi)\sim
O(\varepsilon)$.  The asymptotic limit given by
(\ref{CP_iso_asympt})--(\ref{delta_apr1}) also implies
$\Psi_{\textrm{(II)}}\to 0$ and, even if
$\M_{\textrm{(II)}}$ diverges, we have $\M/x\to\ln(x)/x\to 0$ so that
asymptotically we have also $g_{rr}\to 1$.

\subsubsection{Matching with a vacuum exterior.} 

For a galactic halo structure the cut--off scale for a matching with a
Schwarzschild exterior should be the virial radius $r=\rvir$ (equivalently, $x=c_0$),
hence condition (\ref{darmois32}) implies
\ba\CP_{\textrm{(II)}}(c_0) \ &=& \ 0,\nonumber\\
\quad\Rightarrow\quad
\delta_c \ &=& \
\frac{x_0^2}{6}\left[1+6\,(1+x_0)^4\,(\F(c_0)-\F_0)\right],\nonumber\\
\CP_{\textrm{(II)}} \ &=& \ \F-\F(c_0)\label{delta_sc}\ea
where $\F(c_0)$ is $F(x_0,x)$ evaluated at $x=c_0$.
For $h=0.7$ and bearing in mind that for all virialized galactic halo structures we
have\, $10\,^8\alt\Mvir/M_\odot\alt 10^{15}$,\, numerical values of $c_0$ given by
(\ref{c0}) fall in the range $6 \alt c_0 \alt 30$. Thus, considering the function
$\F$ in (\ref{CP_iso2}) we can expand $\delta_c$ as in (\ref{delta_apr1}), leading to
\begin{equation}\delta_c \ \approx \
\left[\ln\,\frac{1}{\sqrt{x_0}}-0.73\right]\,x_0^2 \ \approx \
x_0^2\,\ln\,\frac{1}{\sqrt{x_0}},\label{delta_apr2}
\end{equation}
so that $\delta_c$ is of the same order of magnitude as in (\ref{delta_apr2}).
Conditions (\ref{darmois31}) for the post--Newtonian metric functions
(\ref{eq_M})--(\ref{grr}) are given by
\ba
\frac{\Phi_c}{c^2}+\varepsilon\,\Psi(c_0)=\varepsilon\,\frac{\M(c_0)}{c_0},\qquad
\M(c_0)=\frac{M(\rvir)}{4\pi\delta_0\rho_0 r_s^3},\label{darmoisPN}\nonumber\\
\ea
which imply
\begin{widetext}
\ba \frac{\Phi_c}{c^2} &=&
-\varepsilon\left[\frac{2+3x_0}{2(1+x_0)^2}+\frac{2}{c_0}\ln\,
\frac{1+c_0}{1+x_0}-\frac{1}{1+c_0}\right]+O(\varepsilon^2) \ \approx \
-\varepsilon\left[1+\frac{2\,\ln(1+c_0)}{c_0}-\frac{1}{1+c_0}\right],\label{Phi_c}
\\ M(\rvir) 
&=& \Mvir\left[1+\frac{3\delta_0}{\Delta
c_0^3}\left(\ln(1+x_0)-\frac{(3+4x_0^2)x_0}{3(1+x_0)^2}\right)+O(\varepsilon)
\right] \ \approx \ \Mvir,\label{M_vir}\ea
\end{widetext}
where we have used the definitions of $\delta_0,\,\Mvir$ in (\ref{delta0}) and
(\ref{rvir}). Notice that, because of the matching with an inner region,
$M(\rvir)=\Mvir$ does not hold exactly in the Newtonian limit, though it holds for a
very good approximation if $x_0$ is sufficiently small. We show in figure 2a the
logarithmic plot of $\CP(x)$ for the two cases considered above (asymptotically
flat and matched to a Schwarzschild exterior).

\subsubsection{Polytropic equation of state.}

The complexity of the expressions in (\ref{CP_iso1}) and (\ref{CP_iso2}) do
not allow us to find out, at first glance, the type of relation between $P$
and $\rho$. Though, by looking at (\ref{Y}) and (\ref{CP_iso_asympt}), the
asymptotic behavior
$\CP\sim\ln(x)/x^4\sim 1/x^4$ and $Y\sim 1/x^3$ indicates a sort of power law relation
between $\CP$ and $Y$ that (at least asymptotically) might be similar to the polytropic
relation (\ref{isotpoly}).  In order to examine the functional relation between $Y$
and $\CP$, we provide in figure 3a the logarithmic plot of $\CP$ vs $Y$ (or
equivalently $\ln P$ vs $\rho\,V_0^2$), for the asymptotically flat case and the
case matched with a Schwarzschild exterior, using the numerical values
$x_0=0.1,\,c_0=8$. For theoretical reference we show the curve corresponding to a
polytropic relation (\ref{isotpoly}) with $n=10$. As shown by the figure, the
asymptotically flat NFW configuration fits very well this polytrope, except for
high density values corresponding to smaller $x$. This behavior is reasonable,
since closer to the center ($x$ close to $x_0$) the NFW density profile becomes
cuspy, while polytropic density profiles are characterized by a ``flat core''. In
the case of a matching with a Schwarzschild exterior, the fitting with a
polytrope also fails near the boundary $x=c_0$ (or $r=\rvir$), which is expected
since we have $\CP(c_0)=0$ while
$Y(c_0)>0$.

\subsection{An anisotropic example}

Since $\CP$ is decoupled from $\M$ and $\Psi$ in the post--Newtonian
field equations (\ref{Mx0})--(\ref{PhiV}), all the expressions for $Y$, \, $\M$ and
$\Psi$ in regions (I) and (II) that we derived in previous sections
(\textit{ie} all equations (\ref{mu_intS})--(\ref{grr}) and
(\ref{darmoisPN})--(\ref{M_vir}), except for (\ref{P_intS}), (\ref{P_intSPN0}) and
(\ref{P_intSPN})) remain valid for the anisotropic case, regardless of the form we
might assume for $\Gamma$. However, equation (\ref{Px0}) does involve $\Gamma$ and so it
must be integrated for both regions. 

A useful expression for the anisotropy factor $\Gamma$ is the ansatz
proposed by Ostipkov and Merritt~\cite{OM}
\begin{equation} \Gamma \ = \ \frac{x^2}{x^2+\beta^2},\label{OM}\end{equation}
where $\beta=r_\beta/r_s=c_0\,r_\beta/\rvir$ marks the length scale (normalized
by $\rvir$) in which the velocities of collisionless particles pass from an isotropic
regime near $x=0$ to a radially dominant mode, since $\Gamma\to 1$ (or $P_\perp\to 0$)
as $x$ becomes larger. Numerical simulations suggest that $P_\perp/P\to 0.6-0.8$ at
about the virial radius $x=c_0$, hence we can set $\beta \sim k\,c_0$ with $1.2\alt k
\alt2$.

Although Darmois matching conditions (\ref{darmois}) allow for jump discontinuities of
$\Gamma$, we will assume the anisotropy factor (\ref{OM}) to be continuous at $x_0$
and to hold also in the domain of the inner region $0\leq x\leq x_0$ with constant
density. Under this assumption, the form equivalent to
$\CP_{\textrm{(I)}}$ in (\ref{CP_iso1}) is
\begin{equation}\CP_{\textrm{(I)}} =
\frac{\beta^2\,\delta_c-\frac{1}{12}\,x^2\,(x^2+2\,\beta^2)}
{x_0^2\,(1+x_0)^4\,(x^2+\beta^2)}+O(\xi),\label{CP_anisI}\end{equation}
while in the region (II) the form of $\CP$ that follows from the integration
of (\ref{Px0}) for (\ref{OM}) and matches continuously with (\ref{CP_anisI}) is
\begin{equation}\CP_{\textrm{(II)}} \ = \
\frac{(x_0^2+\beta^2)\,\CP_0+\Q-\Q_0}{x^2+\beta^2}+O(\varepsilon),
\label{CP_anisII}
\end{equation}
where $\CP_0=\CP_{\textrm{(I)}}(x_0)$,\, $\Q_0=\Q(\beta,x_0)$ with
\begin{widetext}
\ba
\Q(\beta,x_0,x)&=&\frac{\beta_2}{2}\,[\ln(1+x)]^2+[A_2-\beta_2\,\gamma_0]\,
\ln(1+x)+(\beta_2\,\gamma_0-\beta_3)\,\ln(x)+\beta_2\,
\textrm{Li}_2(1+x)-\gamma_0\,B_2-C_2,\nonumber\\
\label{Qdef}\ea
\ba
\beta_2 &=& 1+3\,\beta^2,\quad \beta_3 \ = \ 1+\frac{7}{2}\,\beta^2,\quad \gamma_0 \ =
\
\frac{3(1+x_0)-x_0^2}{3\,(1+x_0)^2}+\ln(1+x_0),\quad
A_2=  
\frac{(7\,\beta^2+2)\,x^3+\beta^2\,(x^2-3x+1)}{2\,x^2\,(1+x)},\nonumber\\
B_2 &=& 
\frac{2\,\beta_2\,x^2+\beta^2\,(3x-1)}{2\,x^2\,(1+x)},
\qquad C_2
=
\frac{(4+13\beta^2)\,x^3+(5+18\beta^2)\,x^2+\beta^2\,(3x-1)}{2\,x^2\,(1+x)^2}\nonumber\ea          
\end{widetext}
Just as in the isotropic case, we examine the asymptotic behavior of the NFW halos
characterized by (\ref{CP_anisI})--(\ref{Qdef}). As mentioned before, the forms
for $\M$ and $\Psi$ and the metric functions given in (\ref{eq_M})--(\ref{grr}) are
valid for these configurations.

\subsubsection{Asymptotically flat cases.}

As oposed to $\CP_{\textrm{(II)}}$ given by (\ref{CP_iso2}),
from (\ref{CP_anisII})--(\ref{Qdef}) we have: $\CP_{\textrm{(II)}}\to 0$ as $x\to
\infty$ for any value we might choose for the parameters $\beta,\,\delta_c$ and $x_0$.
Hence, all NFW configurations characterized by the Ostipkov--Merritt ansatz (\ref{OM})
for $\Gamma$ are asymptotically flat. However, by looking at the asymptotic behavior
of $\Q$
\ba \CP_{\textrm{(II)}} &\approx&
\frac{\C_0}{x^2}
+\frac{\ln\,x-\ln(1+x_0)-2\beta^2\,\C_0-\C_1}{2x^4}\nonumber\\
&&+O\left(\frac{\ln\,x}{x^5}\right)
\label{aniso_asymp}
\\
\C_0&=&(x_0^2+\beta^2)\,\CP_0-\Q_0-\frac{\pi^2}{6}\,\beta_2\qquad
\C_1 = \frac{3-5\,x_0^2}{6\,(1+x_0)^2}\nonumber\ea
it is evident that the asymptotic behavior depends on $\C_0$. If 
$\C_0 >0$, then $\CP_{\textrm{(II)}}>0$ decays asymptotically to zero as $1/x^2$, this
case is shown by unlabeled solid curves in figures 2b and 3b. However, this case is
unphysical because (from(\ref{sigma2})) the velocity dispersion scales asymptotically
as $\sigma^2\to \C_0\,x$ and diverges as $x\to\infty$. If we want
$\sigma^2\to 0$ asymptotically, then we must choose $\C_0 =0$, leading to the same
asymptotic scaling $\CP_{\textrm{(II)}} \sim \ln(x)/x^4$ as in the isotropic case.
From (\ref{CP_anisI}). This case corresponds to the choice

\begin{equation}\delta_c =
\frac{x_0^2}{\beta^2}\,\left[\frac{x_0^2+2\beta}{12}+(1+x_0)^4\,
\left(\Q_0+\beta_2\,\frac{\pi^2}{6}\right)\right],\label{deltac_anis1}\end{equation}
and is marked by the letter {\textbf{A}} in figures 2b and 3b, while
the curves without mark in these figures correspond to various values of
$\C_0>0$.

\subsubsection{Matching with a Schwarzschild exterior}

As in the isotropic case, we assume the matching interface to be $\rvir$ so that
$x=c_0$. The matching conditions (\ref{darmois31}) are given by (\ref{darmoisPN}),
leading also to (\ref{Phi_c}) and (\ref{M_vir}). However, (\ref{darmois32}) in the form
$\CP_{\textrm{(II)}}(c_0)=0$ now implies
\begin{equation}\delta_c \ = \
\frac{x_0^2}{\beta^2}\,\left[\frac{x_0^2+2\,\beta^2}{12}
+(1+x_0)^4\,(\Q_0-\Q(c_0))\right],\label{deltac_anis2}\end{equation}
where $\Q(c_0)$ is given by (\ref{Qdef}) evaluated at $x=c_0$. The form of $\CP$
corresponding to this case is shown as the curve is marked by the letter {\textbf{S}}
in figures 2b and 3b.

\subsubsection{Polytropic equation of state} 

Since $\CP_{\textrm{(II)}}$ with $\C_0=0$ follows the same asymptotic
scaling as in the isotropic case, it is not surprising to find that $\CP$ and $Y$
follow the same approximately polytropic relation. However, in the case
$\C_0> 0$ we see an asymptotic relation of the form $P\sim\rho^{2/3}$ usually for
$x\ll c_0$ far away from the virial radius (see figure 3b).  

\begin{figure}
\centering
% Use the relevant command for your figure-insertion program
% to insert the figure file.
% For example, with the option graphics use
\includegraphics[height=11cm]{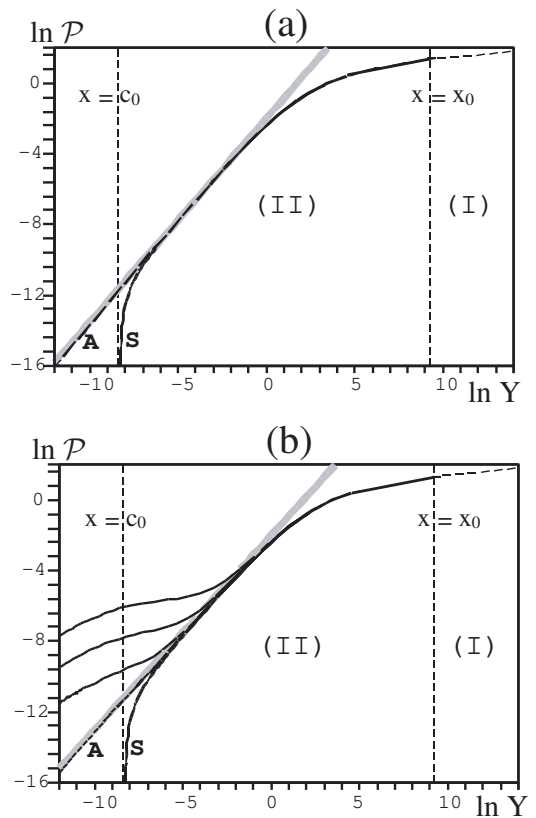}
%
% If not, use
%\picplace{5cm}{2cm} % Give the correct figure height and width in cm
%
\caption{Plot of $\ln\,\CP$ vs $\ln\,Y$, equivalent to plotting $\ln\,P$ vs
$\ln\,\rho\,V_0^2$. Panels (a) and (b) denote the isotropic and anisotropic cases for
the same parameters as in Fig 2. Letters {\textbf {S}} and {\textbf {A}} identify the
curves for the case matched to a Schwarzschild exterior and the case that scales
asymptotically as $(\ln\,x)/x^4$. The other curves in (b) mark the cases that
scale as $x^{-2}$. The dotted curve in the inner region is the ``pure''
NFW without inner region. As a comparison we show a line with slope $1.1$ (thick
grey line) that would correspond to the polytropic relation
$P\propto\rho^{1+1/10}$. Notice how the variables characteristic of the NFW
profile decaying as $(\ln\,x)/x^4$ approximately fit this relation, except near
the center and near the Schwarzschild matching interface.}
\label{fig:3}       % Give a unique label
\end{figure}

\section{Discussion and conclusion}

In the previous sections we have constructed adequate post--Newtonian
generalizations for the galactic halo models that emerge from the well known
NFW numerical simulations. We have shown how the issues of lack of a regular
center (because of interpolating an empiric density profile) and an unbounded halo
mass can be resolved by suitable matchings with a section of an interior
Schwarzschild solution with constant density, and with a vacuum Schwarzschild
exterior. Even if galactic halos are essentially Newtonian systems, we feel it is
important for relativists to see how they can also be described and studied in
General Relativity within the framework of a post--Newtonian weak field regime.
Such a description can be very valuable in studying their interaction with
physical effects (gravitational lenses and gravitational waves) and dark energy
sources, all of which lack an adequate Newtonian description.

Following our proposal that NFW halos satisfy the ideal gas type of equation of
state (\ref{NRIGES}), we have shown empiricaly (see figure 3) that outside their
central core region these halos approximately satisfy the polytropic relation
(\ref{isotpoly}) with $n\approx 10$. This might be quite significant, since
virialized self--gravitating systems are characterized by non--extensive forms
of energy and entropy~\cite{Padma2,Padma3}, and as mentioned before, stellar
polytropes are the equilibrium state associated with the non--extensive entropy
functional in  Tsallis' formalism~\cite{Tsallis,PL,TS1, TS2} (see~\cite{Chavanis}
for a critical appraisal). However,  the consequences of this rough polytropic
relation should be looked carefully, since stellar polytropes are solutions of
Vlassov equation with an isotropic velocity distribution~\cite{BT}, while NFW
halos follow from numerical simulations and exhibit (in general) anisotropic
velocity distributions (even if these anisotropies are not too
large~\cite{LoMa}). In the application of Tsallis formalism to self--gravitating
collisionless systems~\cite{TS1, TS2},  the free parameter
$q=(2n-1)/(2n-3)$ denotes the departure from the extensive Boltzmann--Gibbs entropy
associated with the isothermal sphere (which follows as the limiting case
$n\to\infty$, or equivalently, as $q\to 1$). Assuming Tsallis theory to be
correct, the empiric verification  provided by figure 3 might indicate that in
the region outside  the central core NFW numerical simulations yield
self--gravitating configurations that approach an equilibrium state characterized
by the Tsallis parameter $q\approx 1.1$.

While the central cusps in the density profile predicted by NFW simulations
seem to be at odds with
observations~\cite{cdm_problems_1,cdm_problems_2,vera,LSB}, there is no conflict
between these observations and the
$1/x^3$ scaling of the NFW density profile outside the core region. Although
the issue of the cuspy cores is still controversial, if it turns out that
galactic halos do exhibit flat density cores, their density profiles could 
adjusted to stellar polytropes  and this might be
helpful in providing a better empirical verification of Tsallis' formalism.
However, this idea must be handled with due case, since stellar polytropes
are very idealized configurations. 

Although we have only dealt with NFW halos, the methodology that we have followed
here can be applied, in principle, to any Newtonian model of galactic
halos. For a deeper study of galactic halo models (NFW, as
well as other empiric or theoretical models), it is important to consider 
a wider theoretical framework, not only using a post--Newtonian
approach, but including also the usual thermodynamics of self--gravitation
systems~\cite{Padma2,Padma3}, as well as alternative approaches such as
Tsallis' formalism~\cite{Tsallis,PL,TS1,TS2}. This study might provide
interesting theoretical clues for understanding the Statistical
Mechanics associated with numerical simulations and/or gravitational
clustering. An improvement and extension of the present study of NFW
halos are being pursued elsewhere~\cite{enproceso}.

\section{acknowledgements}

We acknowledges financial support from grants PAPIIT-DGAPA IN--122002 (DN),
PAPIIT-DGAPA number IN--117803 (RAS) and CONACyT 32138--E and 34407--E (TM).

\end{document}